\documentclass[seceq]{ptptex}
\usepackage{epsfig}

\title{Chiral symmetry, strangeness and resonances}

\author{M.F.M. \textsc{Lutz}$^{a,b}$,  E.E. \textsc{Kolomeitsev}$^c$ and C.L. \textsc{Korpa}$^d$ }

\inst{$^a$Gesellschaft f\"ur Schwerionenforschung GSI, Postfach 110552\\
D-64220 Darmstadt, Germany \\
$^b$Institut f\"ur Kernphysik, TU Darmstadt\\
D-64289 Darmstadt, Germany \\
$^c$The Niels Bohr Institute, Blegdamsvej 17\\ DK-2100 Copenhagen, Denmark \\
$^d$Department of Theoretical Physics, University of
Pecs, \\Ifjusag u.\ 6, 7624 Pecs, Hungary}

\abst{ We review the important role played by the chiral SU(3)
symmetry in predicting the properties of antikaons and hyperon resonances
in cold nuclear matter. Objects of
crucial importance are the meson-baryon scattering amplitudes
obtained within the chiral coupled-channel effective field theory.
The formation of baryon resonances as implied by chiral coupled-channel dynamics
is discussed. Results for antikaon and
hyperon-resonance spectral functions are
presented for isospin symmetric and asymmetric matter. }

\def\half{{\textstyle{1\over 2}}}

\def\prt{\partial}

\begin{document}

\maketitle

\section{Introduction}

In this talk we review the application of chiral
coupled-channel dynamics to nuclear matter properties of the
antikaon, the $\Lambda (1405)$ s-wave resonance and the
$\Sigma(1385)$ p-wave resonance. A softening of the antikaon mode in nuclear
could have been already
anticipated in the 70's from K-matrix analysis of
the antikaon-nucleon scattering data (see e.g. \cite{ADMartin}) which
predicted considerable attraction in the subthreshold s-wave $K^-$
nucleon scattering amplitudes. In conjunction with the low-density
theorem \cite{dover,njl-lutz} this leads to an antikaon
spectral function in nuclear matter that has significant strength at energies
smaller than the kaon mass. As was pointed out first in
\cite{ml-sp} the realistic evaluation of the antikaon self energy
in nuclear matter requires a self consistent scheme. The feedback
effect of the antikaon spectral function on the
antikaon-nucleon scattering process was found to be important for
the hyperon resonance structure in nuclear matter. In turn the
modified structure of the hyperon resonances influences the spectral function of the
antikaon.

We present and discuss up-to-date results for the spectral functions of antikaons
in symmetric and asymmetric cold nuclear matter that are based on a chiral-coupled
channel analysis of meson-baryon scattering data \cite{LK02,Lutz:Korpa,Korpa:Lutz:03}.
Unfortunately the empirical data set still leaves much room for different
theoretical interpretations, in particular for the subthreshold scattering amplitudes that
determine the amount of attraction an antikaon is subject to in cold nuclear matter.
Until recently acceptable effective field theories were  much less developed
in the strangeness sectors as compared to the strangeness zero sector.
As a consequence different analysis lead to scattering amplitudes that
differ by about a factor two at subthreshold energies ( see \cite{Florkowski,Cieply}).
Thus, it is useful to review also in detail effective coupled-channel field theories
based on the chiral Lagrangian.

The task to construct a systematic effective field theory for the
meson-baryon scattering processes in the resonance region is
closely linked to the fundamental question as to what is the
'nature' of baryon resonances. The radical conjecture
\cite{LK01,LK02,LWF02,LH02} that meson and baryon resonances not belonging
to the large-$N_c$ ground states are generated by
coupled-channel dynamics lead to a series of works
\cite{Granada,KL03,LK03,LK03-2,KL03-2,HL03}
demonstrating the crucial importance of coupled-channel dynamics for resonance physics in QCD.
This conjecture was  challenged by a phenomenological model
\cite{LWF02}, which generated successfully non-strange s- and
d-wave resonances by coupled-channel dynamics describing a large
body of pion and photon scattering data. Of course, the idea to
explain resonances in terms of coupled-channel dynamics is an old one going back to
the 60's \cite{Wyld,Dalitz,Ball,Rajasekaran,Wyld2,sw88}. For a
comprehensive discussion of this issue we refer to \cite{LH02}.
In recent works \cite{Granada,KL03}, which will be reviewed here,
it was shown that chiral dynamics as implemented by the
$\chi-$BS(3) approach \cite{LK00,LK01,LK02,LH02} provides a
parameter-free leading-order prediction for the existence of a wealth of strange
and non-strange s- and d-wave wave baryon resonances. A quantitative description
of the low-energy pion-, kaon and antikaon scattering data was achieved earlier
within the $\chi$-BS(3) scheme upon incorporating chiral correction terms \cite{LK02}.

\section{Effective field theory of chiral coupled-channel dynamics}

\begin{figure}[b]
\begin{center}
\includegraphics[width=12.0cm,clip=true]{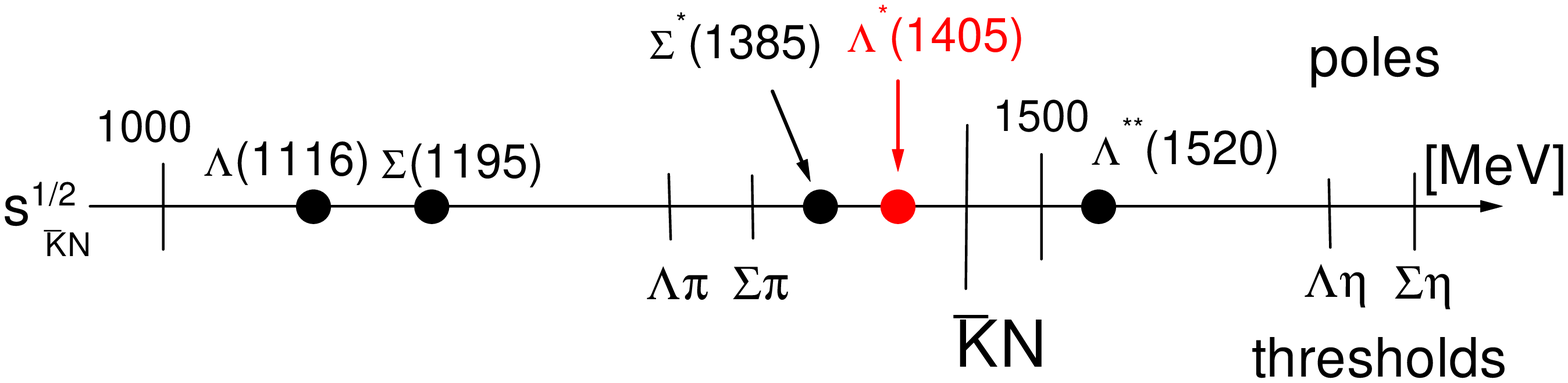}
\end{center}
\caption{The world of antikaon-nucleon scattering} \label{fig1}
\end{figure}

Consider for instance the rich world of antikaon-nucleon
scattering illustrated in Fig. \ref{fig1}. The figure clearly
illustrates the complexity of the problem. The $\bar K N$ state
couples to various inelastic channel like $\pi \Sigma$ and $\pi
\Lambda$, but also to baryon resonances below and above its
threshold. The goal is to bring order into this world seeking a
description of it based on the symmetries of QCD. For instance, as
will be detailed below, the $\Lambda(1405)$ and $\Lambda(1520)$
resonances will be generated by coupled-channel dynamics, whereas
the $\Sigma(1385)$ should be considered as a 'fundamental' degree of
freedom. Like the nucleon and hyperon ground states the
$\Sigma(1385)$ enters as an explicit field in the effective
Lagrangian set up to describe the $\bar K N$ system.

The starting point to describe the meson-baryon scattering process
is the chiral SU(3) Lagrangian (see e.g.\cite{Krause,LK02}). A
systematic approximation scheme arises due to a successful scale
separation justifying the chiral power counting rules
\cite{book:Weinberg}. The effective field theory of the
meson-baryon scattering processes is based on the assumption that
the scattering amplitudes are perturbative at subthreshold
energies with the expansion parameter $Q/ \Lambda_{\chi}$. The
small scale $Q$ is to be identified with any small momentum of the
system. The chiral symmetry breaking scale is
$$\Lambda_\chi \simeq 4\pi f \simeq 1.13 \;{\mbox GeV}\,, $$ with the parameter $f\simeq 90$ MeV
determined by the pion decay process. Once the available energy is
sufficiently high to permit elastic two-body scattering a further
typical dimensionless parameter $m_K^2/(8\,\pi f^2) \sim 1$ arises
\cite{LK00,LK01,LK02}. Since this ratio is uniquely linked to
two-particle reducible diagrams it is sufficient to sum those
diagrams keeping the perturbative expansion of all irreducible
diagrams, i.e. the  coupled-channel Bethe-Salpeter equation has to
be solved. This is the basis of the $\chi$-BS(3) approach
developed in \cite{Lutz00,NA00,LK00,LK01,LK02}.

At leading order in the chiral expansion one encounters the famous
Weinberg-Tomozawa \cite{Wein-Tomo} interaction,
\begin{eqnarray}
\mathcal{L}_{WT}&=&
 \frac{i}{8\, f^2}\, {\rm tr}\, \Big((\bar B
\,\gamma^\mu\,\, B) \cdot
 [\Phi,(\prt_\mu \Phi)]_-  \Big)
\nonumber\\
&+&\frac{3\,i}{8\, f^2}\, {\rm tr}\, \Big( (\bar B_\nu
\,\gamma^\mu\,\, B^\nu) \cdot
 [\Phi,(\prt_\mu \Phi)]_-  \Big)
 \,,
 \label{WT-term}
\end{eqnarray}
where we dropped additional structures that do not contribute to
the on-shell scattering process at tree level. The terms in
(\ref{WT-term}) constitute the leading order s-wave interaction of
Goldstone bosons ($\Phi $) with the baryon-octet ($B$) and
baryon-decuplet ($B_\mu$) states. The octet and decuplet fields,
$\Phi, B$ and $B_\mu$, posses an appropriate matrix structure
according to their SU(3) tensor representation.

The scattering process is described by the amplitudes that follow
as solutions of the Bethe-Salpeter equation,
\begin{eqnarray}
T(\bar k ,k ;w ) &=& K(\bar k ,k ;w ) +\int\!\!
\frac{d^4l}{(2\pi)^4}\,K(\bar k , l;w )\, G(l;w)\,T(l,k;w )\;,
\nonumber\\
G(l;w)&=&-i\,D(\half\,w-l)\,S( \half\,w+l)\,, \label{BS-coupled}
\end{eqnarray}
where we suppress the coupled-channel structure for simplicity.
The meson and baryon propagators,  $D(q)$ and $S(p)$, are used in
the notation of \cite{LWF02}. We apply the convenient kinematics:
\begin{eqnarray}
w = p+q = \bar p+\bar q\,, \quad k= \half\,(p-q)\,,\quad \bar k
=\half\,(\bar p-\bar q)\,, \label{def-moment}
\end{eqnarray}
where $q,\,p,\, \bar q, \,\bar p$ the initial and final meson and
baryon 4-momenta. The Bethe-Salpeter scattering equation is
recalled for the case of meson baryon-octet scattering. An
analogous equation holds for meson baryon-decuplet scattering
process (see e.g. \cite{LWF02}).

The scattering amplitude $T(\bar k,k;w)$ decouples into various
sectors characterized by isospin ($I$) and strangeness ($S$)
quantum numbers. In the case of meson baryon-octet and
baryon-decuplet scattering the following channels are relevant
\begin{eqnarray}
&& (I,S)_{[8 \otimes 8]} =  (0,-3), (1,-3), (\frac{1}{2},-2),
(\frac{3}{2},-2) ,
(0,-1),  \nonumber\\
&& \qquad \qquad (1,-1), (2,-1), (\frac{1}{2},0),(\frac{3}{2},0),
(0,1),(1,1) \,,
\nonumber\\
&& (I,S)_{[8 \otimes 10]} = (\frac{1}{2},-4), (0,-3), (1,-3),
(\frac{1}{2},-2), (\frac{3}{2},-2) ,
(0,-1), \nonumber\\
&& \qquad \qquad  (1,-1), (2,-1),(\frac{1}{2},0),(\frac{3}{2},0),
(\frac{5}{2},0), (1,1),(2,1) \,. \label{sectors-10}
\end{eqnarray}

Referring to the detailed discussion given in \cite{LK02,LK03} we
assume a systematic on-shell reduction of the Bethe-Salpeter
interaction kernel. We introduce an on-shell equivalent effective
interaction kernel $V$, together with three off-shell interaction
kernels $V_L,V_R$ and $V_{LR}$ where $V_R$ ($V_L$) vanishes if the
initial (final) particles are on-shell. The interaction kernel
$V_{LR}$ is defined to vanish if evaluated with either initial or
final particles on-shell. The latter objects are defined by:
\begin{eqnarray}
K &=&V+ (1-V \cdot G)\cdot V_{L}+ V_{R}\cdot (1-G\cdot V)\,
\nonumber\\
&+& (1-V\cdot G)\cdot V_{LR}\cdot (1-G\cdot V) - V_R\cdot
\frac{1}{1-G\cdot V_{LR}}\cdot G \cdot V_L \,. \label{K-decomp}
\end{eqnarray}
The decomposition of the Bethe-Salpeter interaction kernel is
unique and can be applied to an arbitrary interaction kernel once
it is defined what is meant with the 'on-shell' part of any
two-particle amplitude. The latter we define as the part of the
amplitude that has a decomposition into the complete set of
projectors
\begin{eqnarray}
V(\bar k,k;w) = \sum_{J, P,a,b}\,V^{(J P)}_{ab}(\sqrt{s})\,
{\mathcal Y}^{(J P)}_{ab}(\bar q,q;w) \,, \label{v-exp}
\end{eqnarray}
where the projectors carry good total angular momentum $J$ and
parity $P$. The merit of the projectors is that they decouple the
Bethe-Salpeter equation (\ref{BS-coupled}) into orthogonal sectors
labeled by the total angular momentum, $J$, and parity, $P$. We
emphasize that the projectors have also the important property
that they are applicable in the case of intermediate states that
have broad spectral distributions. It is clear
that performing a chiral expansion of $K$ and $V$ to some order
$Q^n$ leads to a straight forward identification of the off-shell
kernels $V_L,V_R$ and $V_{LR}$ to the same accuracy.

The on-shell part of the scattering amplitude takes the simple
form,
\begin{eqnarray}
&& T^{\rm on-shell}(\bar k ,k ;w )  = \sum_{J,P}\,M^{(J
P)}(\sqrt{s}\,)\, {\mathcal Y}^{(J P)}(\bar q, q;w) \,,
\nonumber\\
&& M^{(J P)}(\sqrt{s}\,) = \Big[ 1- V^{(J P)}(\sqrt{s}\,)\,J^{(J
P)}(\sqrt{s}\,)\Big]^{-1}\, V^{(J P)}(\sqrt{s}\,)\,, \label{}
\end{eqnarray}
with a set of divergent loop functions $J^{(J P)}(\sqrt{s}\,)$. We
insist on the renormalization condition,
\begin{eqnarray}
T^{(I,S)}(\bar k,k;w)\Big|_{\sqrt{s}= \mu (I,S)} = V^{(I,S)}(\bar
k,k;w)\Big|_{\sqrt{s}= \mu (I,S)} \,, \label{ren-cond}
\end{eqnarray}
together with the natural choice for the subtraction points,
\begin{eqnarray}
&& \mu(I,+1)=\mu(I,-3)={\textstyle{1\over 2}}\,(m_\Lambda+
m_\Sigma) \,, \quad \mu(I,0)=m_N\,, \quad
\nonumber\\
&&  \mu(0,-1)=m_\Lambda,\quad \mu(1,-1)=m_\Sigma\,, \quad
\mu(I,-2)= \mu(I,-4)= m_\Xi \,,
 \label{eq:sub-choice}
\end{eqnarray}
as explained in detail in \cite{LK02,Granada,KL03}. The renormalization
condition reflects the basic assumption our effective field theory
is based on, namely that at subthreshold energies the scattering
amplitudes can be evaluated in standard chiral perturbation
theory. This is achieved by supplementing (\ref{BS-coupled}) with
(\ref{ren-cond},\ref{eq:sub-choice}).  The subtraction points
(\ref{eq:sub-choice}) are the unique choices that protect the
s-channel baryon-octet masses manifestly in the p-wave
$J={\textstyle{1\over 2}}$ scattering amplitudes.

It is useful to elaborate  in some detail on the structure of the
loop functions. The merit of the projector technique is that
dimensional regularization can be used to evaluate the latter
ones. Here we exploit the result that any given projector is a
finite polynomial in the available 4-momenta. This implies that
the loop functions can be expressed in terms of a log-divergent
master function, $I(\sqrt{s}\,)$, and reduced tadpole terms,
\begin{eqnarray}
&& J^{(J P)}(\sqrt{s}\,)= N^{(J
P)}(\sqrt{s}\,)\,\Big[I(\sqrt{s}\,) -I(\mu )\Big] \,,
\nonumber\\
&& I(\sqrt{s}\,)=\frac{1}{16\,\pi^2} \left(
\frac{p_{cm}}{\sqrt{s}}\, \left( \ln
\left(1-\frac{s-2\,p_{cm}\,\sqrt{s}}{m^2+M^2} \right) -\ln
\left(1-\frac{s+2\,p_{cm}\sqrt{s}}{m^2+M^2} \right)\right) \right.
\nonumber\\
&&\qquad \qquad + \left.
\left(\frac{1}{2}\,\frac{m^2+M^2}{m^2-M^2} -\frac{m^2-M^2}{2\,s}
\right) \,\ln \left( \frac{m^2}{M^2}\right) +1 \right)+I(0)\;,
\label{i-def}
\end{eqnarray}
where $\sqrt{s}= \sqrt{M^2+p_{cm}^2}+ \sqrt{m^2+p_{cm}^2}$. The
normalization factor $N_a^{(J P)}(\sqrt{s}\,)$ is a polynomial in
$\sqrt{s}$ and the mass parameters. In (\ref{i-def}) the
renormalization scale dependence of the scaler loop function
$I(\sqrt{s}\,)$ was traded in favor of a dependence on a
subtraction point $\mu$, leading to compliance with the
renormalization condition (\ref{ren-cond}). The loop functions
$J^{(J,P)}(\sqrt{s}\,)$ are consistent with chiral counting rules
only if the subtraction scale $\mu \simeq M$ is chosen close to
the 'heavy' hadron mass \cite{LK00,LK02}. Moreover it was shown
that keeping reduced tadpole terms in the loop functions leads to
a renormalization of s-channel exchange terms that is in conflict
with chiral counting rules if the effective interaction kernel is evaluated
in perturbation theory \cite{LK02}.

\begin{figure}[t]
\begin{center}
\includegraphics[width=9.0cm,clip=true]{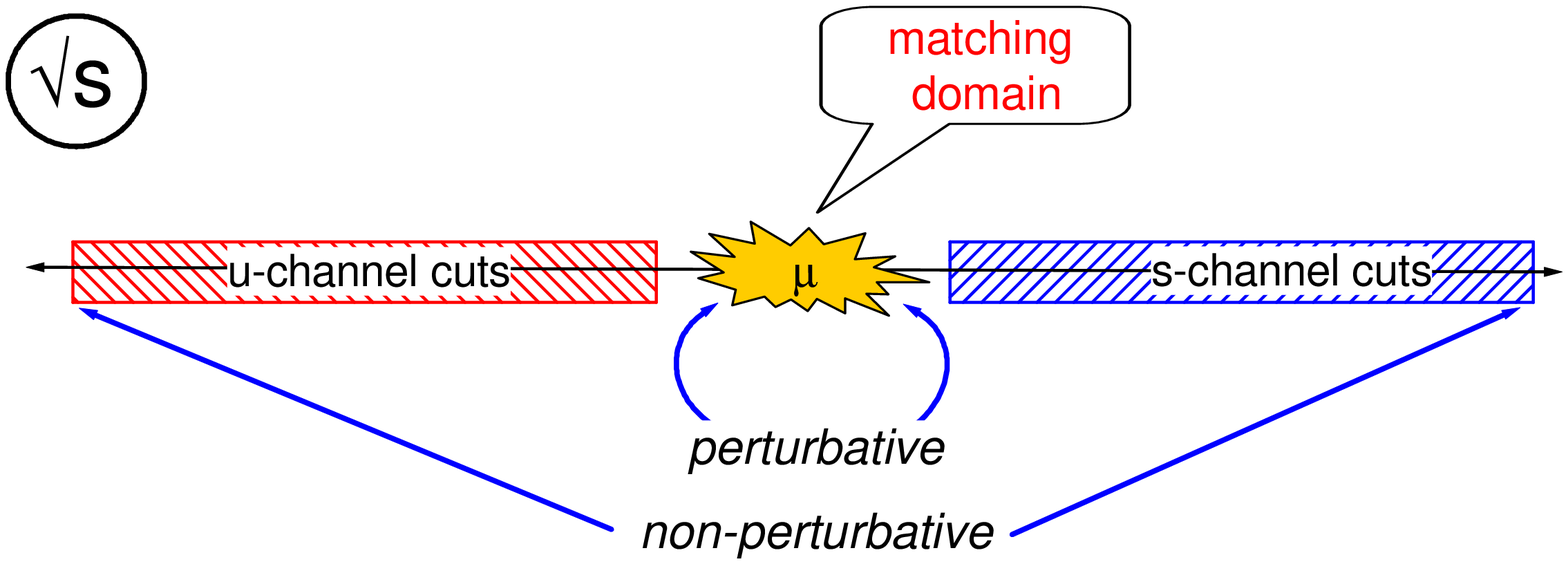}
\end{center}
\caption{Graphical illustration for the gluing of s- and u-channel
unitarized scattering amplitudes} \label{fig4}
\end{figure}

The merit of the scheme \cite{LK00,LK01,LK02} lies in the property
that for instance the $K \,\Xi$ and $\bar K\,\Xi $ scattering
amplitudes match at $\sqrt{s} \sim m_\Xi $ approximately as
expected from crossing symmetry. In \cite{LK02,LK03} we suggested to
glue s- and u-channel unitarized scattering amplitudes at
subthreshold energies as illustrated in Fig. \ref{fig4}.
This reflects the basic assumption that diagrams
showing an s-channel or u-channel unitarity cut need to be summed
to all orders at least at energies close to where the diagrams
develop their imaginary part. By construction, a glued scattering
amplitude satisfies crossing symmetry exactly at energies where
the scattering process takes place. At subthreshold energies
crossing symmetry is implemented approximatively only, however, to
higher and higher accuracy when more chiral correction terms are
considered. Insisting on the renormalization condition
(\ref{ren-cond},\ref{eq:sub-choice}) guarantees that subthreshold
amplitudes match smoothly and therefore the final 'glued'
amplitudes comply with the crossing-symmetry constraint to high
accuracy. Fig. \ref{fig4} illustrates this mechanism  for two typical cases \cite{LK03}.
Whereas in case I) both the s- and u-channel unitarizations lead to bound states,
in case II) only the u-channel unitarization generates a bound state. For simplicity
the figure does not show the additional complications
present in the meson-baryon forward scattering amplitude, that are
due to pole structures implied by the baryon octet ground states
\cite{LK02}. It is amusing to observe that the natural subtraction
points (\ref{eq:sub-choice}) can also be derived if one incorporates photon-baryon inelastic
channels. Then additional constraints arise. For instance the
reaction $\gamma \,\Xi \to \gamma \,\Xi $, which is subject to a
crossing symmetry constraint at threshold, may go via the
intermediate states $\bar K \,\Lambda $ or $\bar K \,\Sigma $.

\begin{figure}[t]
\begin{center}
\includegraphics[width=9.0cm,clip=true]{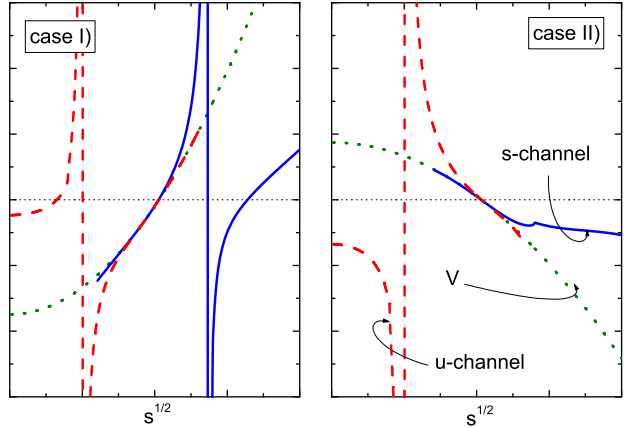}
\end{center}
\vspace*{-0.3cm}
\caption{Illustration of the gluing of s- and u-channel unitarized
scattering amplitudes for two typical cases. The lines extending
to the right (left) are the results of a s-channel (u-channel)
unitarization. The dotted lines represent the contribution of the effective interaction kernel $V$}
 \label{fig4}
\end{figure}

The perturbative nature of subthreshold amplitudes, a crucial
assumption of the $\chi$-BS(3) approach proposed in
\cite{LK00,LK01,LK02}, is not necessarily true in phenomenological
coupled-channel schemes in
\cite{ksw95,grnpi,grkl,Oset-prl,Oset-plb,Jido03}. Using the
subtraction scales as free parameters, as advocated in
\cite{Oset-prl,Oset-plb,Jido03}, may be viewed as promoting the
counter terms of chiral order $Q^3$ to be unnaturally large. If
the subtraction scales are chosen far away from their natural
values (\ref{eq:sub-choice}) the resulting loop functions are in
conflict with chiral power counting rules \cite{LK00}. Though
unnaturally large $Q^3$ counter terms can not be excluded from
first principals one should check such an assumption by studying
corrections terms systematically. A detailed test of the
naturalness of the $Q^3$ counter terms was performed within the
$\chi$-BS(3) scheme \cite{LK02} demonstrating good convergence in
the channels studied without any need for promoting the counter
terms of order $Q^3$. Possible correction terms in the approach
followed in \cite{Oset-prl,Oset-plb,Jido03} have so far not been
studied systematically for meson-baryon scattering. Moreover, if
the scheme advocated in \cite{Oset-prl,Oset-plb,Jido03} were
applied in all eleven isospin strangeness sectors with
$J^P=\frac{1}{2}^-$  a total number of 26 subtraction parameters
arise. This should be compared with the only ten counter terms of
chiral order $Q^3$ contributing to the on-shell scattering
amplitude at that order \cite{LK02}. Selecting only the operators
that are leading in the large-$N_c$ limit of QCD out of the ten
$Q^3$ operators only four survive \cite{LK02}. We conclude that it
would be inconsistent to apply the approach used in
\cite{Oset-prl,Oset-plb,Jido03} in all isospin strangeness
channels without addressing the above mismatch of parameters. Our
scheme has the advantage over the one
in~\cite{Oset-prl,Oset-plb,Jido03} that once the parameters
describing subleading effects are determined in a subset of
sectors one has immediate predictions for all sectors $(I,S)$. A
mismatch of the number of parameters is avoided altogether since
the $Q^3$ counter terms enter the effective interaction kernel
directly.

\begin{figure}[t]
\begin{center}
\includegraphics[width=9.0cm,clip=true]{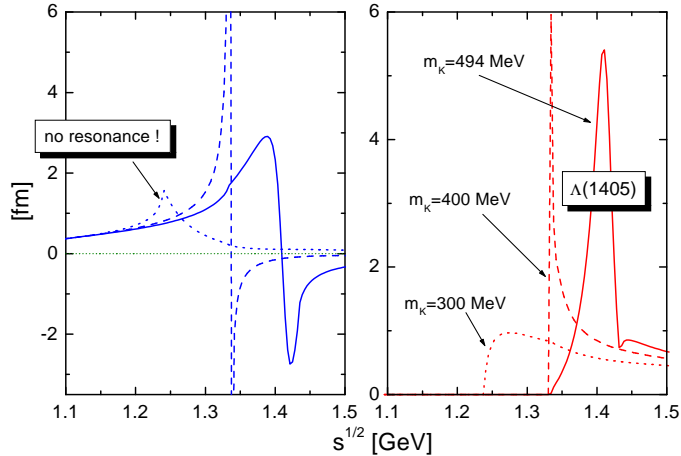}
\end{center}
\caption{S-wave antikaon-nucleon scattering amplitude in the isospin zero channel as it
follows in the $\chi$-BS(3) approach at leading order. } \label{fig5}
\end{figure}

Given the subtraction scales (\ref{eq:sub-choice}) the leading
order calculation is parameter free. Of course chiral correction
terms do lead to further so far unknown parameters  which need to
be adjusted to data. Within the $\chi-$BS(3) approach such
correction terms enter the effective interaction kernel $V$ rather
than leading to subtraction scales different from
(\ref{eq:sub-choice}) as it is assumed
in~\cite{Oset-prl,Oset-plb,Jido03}. In particular the leading
correction effects are determined by the counter terms of chiral
order $Q^2$.

The $\chi$-BS(3) scheme was applied first in \cite{LK00,LK01} where results for the
antikaon-nucleon scattering amplitudes at leading and subleading orders were presented.
In particular it was demonstrated that the $\Lambda(1405)$ resonance is generated by coupled-channel
dynamics without the need of adjusting any parameter \cite{LK00}. This is an important result because it
avoided for the first time the use of a fine-tuned cutoff parameter which is at odds with
chiral counting rules. The isospin zero s-wave scattering amplitude as it followed
at leading order is shown in Fig. \ref{fig4}, which  clearly illustrates the presence of the
$\Lambda(1405)$ resonance. Lowering the
strange current quark mass of QCD has a dramatic effect on the resonance structure. Chiral-coupled
channel dynamics predicts that in a world with kaons of mass 300 MeV the $\Lambda(1405)$ resonance
would not exist. This is a clear prediction that can be tested with Lattice QCD simulations
\cite{lattice:1,lattice:2,lattice:3,dgr}.
Moreover the figure reflects the matching of the unitarized scattering amplitude with the
effective interaction kernel at subthreshold energies. As a consequence of the defining assumption of
the $\chi$-BS(3) approach, namely, that the subthreshold scattering amplitudes remain perturbative,
all lines in the figure representing the real part of the scattering amplitude join at subthreshold
energies. This follows since the effective interaction kernel does not depend on the kaon mass at
leading order \cite{LK02}.

\section{Baryon resonances from chiral SU(3) symmetry}

There is a long standing controversy to what is the nature of
s-wave baryon resonances. Before the event of the quark model
several such states have been successfully generated in terms of
coupled-channels dynamics
\cite{Wyld,Dalitz,Ball,Rajasekaran,Wyld2}. A prime example is the $\Lambda(1405)$ resonance,
discussed already in the previous section, and
which was successfully described already  in the latter works.
These early calculations are closely related to modern approaches based on the
leading-order chiral SU(3) Lagrangian. The interaction used in
\cite{Wyld,Dalitz,Ball,Rajasekaran,Wyld2} matches the
Weinberg-Tomozawa interaction (\ref{WT-term}) if expanded in a
Taylor series \cite{sw88}. The main difference of the early
attempts from computations based on the chiral Lagrangian is the
way the coupled-channel scattering equation is regularized and
renormalized. The crucial advance over the last years in this
field is therefore a significant improvement of the systematics,
i.e. how to implement corrections terms into coupled-channel
dynamics.

We give a discussion of the s- and d-wave baryon resonance
spectrum that arises in chiral-coupled effective field theory based on the
leading order chiral Lagrangian \cite{Granada,KL03}. Consider first the
the SU(3) limit. The latter is not defined
uniquely depending on the magnitude of the current quark masses,
$m_u=m_d=m_s$. We study two scenarios \cite{Granada,KL03}. In
the 'light' SU(3) limit the current quark masses are chosen such
that one obtains $m_\pi= m_K=m_\eta =140$ MeV. The second case,
the 'heavy' SU(3) limit, is characterized by $m_\pi= m_K=m_\eta
=500$ MeV.

The $J^P=\frac{1}{2}^-$ baryon resonances manifest themselves as poles in the
s-wave  meson baryon-octet scattering amplitudes. In the SU(3) limit the latter
decompose according to,
\begin{eqnarray}
&& 8 \otimes 8 = 27 \oplus \overline{10} \oplus 10 \oplus 8 \oplus
8 \oplus 1 \, \label{8-8-decom}
\end{eqnarray}
The leading order chiral Lagrangian (\ref{WT-term}) predicts attraction in the two
octet and the singlet channel but repulsion in the 27-plet and
decuplet channels. In the 'heavy' SU(3) limit the
chiral dynamics predicts two degenerate octet bound states
together with a non-degenerate singlet state
\cite{Wyld,Dalitz,Ball,Rajasekaran,Wyld2,Granada,Jido03,Granada,KL03}.
However, in the 'light' SU(3) limit all states disappear leaving no clear
resonance signal.

In the $J^P=\frac{3}{2}^-$ sector the resonances properties are studied in terms of
the s-wave meson baryon-decuplet scattering amplitudes. In this case the
Weinberg-Tomozawa interaction (\ref{WT-term}) is attractive in the
octet, decuplet and 27-plet channel, but repulsive in the 35-plet
channel,
\begin{eqnarray}
&&8 \otimes 10 = 35 \oplus 27 \oplus 10 \oplus 8\,.
\label{8-10-decom}
\end{eqnarray}
Therefore one may expect resonances or bound states in the former
channels. Indeed, in the 'heavy' SU(3) limit $72=4\times
(8+10)$ bound states are generated in this sector forming an octet and decuplet
representation of the SU(3) group. A 27-plet-bound
state is not observed. This reflects the weaker attraction in that channel. However,
if one artificially increases the amount of attraction by about 40
$\%$ by lowering the value of $f$ in the Weinberg-Tomozawa term, a
clear bound state arises in this channel also. A contrasted result
is obtained if one lowers the meson masses down to the pion mass
arriving at the 'light' SU(3) limit. Then we find neither bound
nor resonance octet or decuplet states. This pattern is a clear
prediction of chiral couple-channel dynamics which should be
tested with unquenched QCD lattice simulations \cite{lattice:1,lattice:2,lattice:3,dgr}.

\begin{figure}[t]
\begin{center}
\includegraphics[width=10.0cm,clip=true]{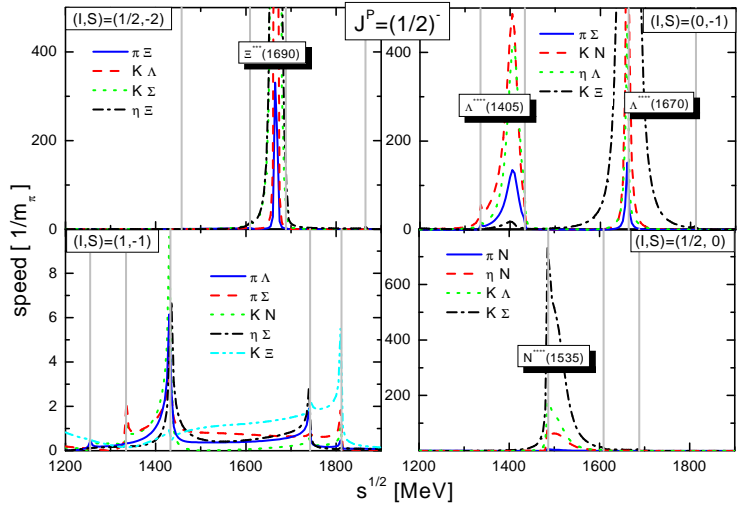}
\end{center}
\caption{Diagonal speed plots of the $J^P=\frac{1}{2}^-$ sector.
The vertical lines show the opening of inelastic meson
baryon-decuplet channels. Parameter-free results are obtained in
terms of physical masses and $f=90$ MeV \cite{Granada,KL03}.}
\label{fig6}
\end{figure}

\begin{figure}[t]
\begin{center}
\includegraphics[width=10.0cm,clip=true]{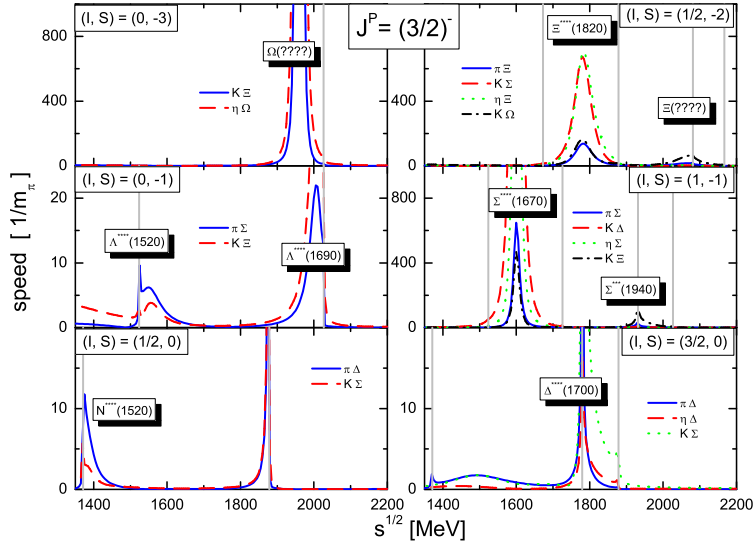}
\end{center}
\caption{Diagonal speed plots of the $J^P=\frac{3}{2}^-$ sector.
The vertical lines show the opening of inelastic meson
baryon-decuplet channels. Parameter-free results are obtained in
terms of physical masses and $f=90$ MeV.} \label{fig7}
\end{figure}

Using physical meson and baryon masses the bound-state turn into
resonances as shown in Figs. \ref{fig6},\ref{fig7} in terms of
speed plots. In \cite{KL03}
we generalized the notion of a speed \cite{Hoehler:speed} to the
case of coupled-channels in a way that the latter reveals the
coupling strength of a given resonance to any channel, closed or
open. The merit of producing speed plot lies in a
convenient property of the latter allowing a straight forward
extraction of resonance parameters. Assume that a coupled-channel
amplitude $M_{ab}(\sqrt{s}\,)$ develops a pole of mass $m_R$, with
\begin{eqnarray}
&& M_{ab}(\sqrt{s}\,) = -\frac{
g^*_a\,g^{\phantom{*}}_b\,m_R}{\sqrt{s}-m_R + i\,\Gamma/2}
\,,\qquad \Gamma_a =
\frac{|g_a|^2}{4\,\pi}\,|p^{(a)}_{cm}|\,N_a(m_R)\,,
\label{def-res}
\end{eqnarray}
where the total resonance width, $\Gamma$, is given by the sum of
all partial widths. The normalization factor $N (M_R)$ in
(\ref{def-res}) is identical to the one entering the form of the
loop functions in (\ref{i-def}). The speed plots take a maximum at
the resonance mass $\sqrt{s}=m_R$, with
\begin{eqnarray}
&& {\rm Speed}_{aa}(m_R\,) = \Bigg\{
\begin{array}{ll}
2\,\frac{\Gamma_a}{\Gamma^2}\,\Big|2\,\sum_{c}\,\frac{\Gamma_c}{\Gamma}
-1\Big| \,
\qquad \qquad  \,{\rm if} \; a= {\rm open}\\
2\,\frac{\Gamma_a}{\Gamma^2}\,\Big|2\,
\sum_{c}\,\frac{\Gamma_c}{\Gamma} -i\,\Big| \, \qquad \qquad
\,{\rm if} \; a= {\rm closed} \,,
\end{array}
\nonumber\\
&& \Gamma = \sum_{a={\rm open}} \Gamma_a \,. \label{speed-an}
\end{eqnarray}
The result (\ref{speed-an}) clearly demonstrates that the speed of
a resonance in a given open channel $a$ is not only a function of
the total width parameter $\Gamma$ and the partial width
$\Gamma_a$. It does depend also on how strongly closed channels
couple to that resonance. This is in contrast to the delay time
of a resonance for which closed channels do not
contribute. In the case of s-wave resonances thresholds induce
square-root singularities which should not be confused with a resonance
signal.

The speed plots of Fig. \ref{fig6} show evidence for the formation
of the $\Xi(1690)$, $\Lambda(1405)$, $\Lambda(1670)$ and $N(1535)$
resonances close to their empirical masses. An additional
$(I,S)=(0,-1)$ state, which couples strongly to the $\pi \Sigma$ channel
\cite{Jido03,Granada}, can be found as a complex pole in the
scattering amplitude close to the pole implied by the
$\Lambda(1405)$ resonance. There is no clear signal for
$(I,S)=(1,-1)$ resonances at this leading order calculation.
However, chiral corrections lead to a clear signal in this sector
\cite{LK02} suggesting a state that may be identified with the
$\Sigma(1750)$ resonance, the only well established s-wave
resonance in this sector. The fact that a second resonance with
$(I,S)=(\frac{1}{2},0)$ is not seen in Fig. \ref{fig6}, even
though the 'heavy' SU(3) limit suggests its existence, may be taken as
a confirmation of the phenomenological observation \cite{LWF02}
that the $N(1650)$ resonance couples strongly to the $\omega_\mu
\,N$ channel not considered here.

In Fig. \ref{fig7} speed plots of the $J^P=\frac{3}{2}^-$ sector
are shown for all channels in which octet and decuplet resonance
states are expected. It is a remarkable success of the
$\chi$-BS(3) approach that it predicts the four star hyperon
resonances $\Xi(1820)$, $\Lambda(1520)$, $\Sigma (1670)$ with
masses quite close to the empirical values. The nucleon and isobar
resonances $N(1520)$ and $\Delta (1700)$ also present in Fig.
\ref{fig7}, are predicted with less accuracy. The important result
here is the fact that those resonances are generated at all. It
should not be expected to obtain already fully realistic results
in this leading order calculation. For instance chiral correction
terms are expected to provide a d-wave $\pi \,\Delta$-component of
the $N(1520)$. We continue with the peak in the (0,-3)-speeds at
mass 1950 MeV. Since this is below all thresholds it is in fact a
bound state. Such a state has so far not been observed but is
associated with a decuplet resonance \cite{Schat}. Further states
belonging to the decuplet are seen in the $(\frac{1}{2},-2)$- and
$(1,-1)$-speeds at masses 2100 MeV and 1920 MeV. The latter state
can be identified with the three star $\Xi (1940)$ resonance.
Finally we point at the fact that the $(0,-1)$-speeds show signals
of two resonance states consistent with the existence of the four
star resonance $\Lambda(1520)$ and $\Lambda(1690)$ even though in
the 'heavy' SU(3) limit we observed only one bound state. It
appears that the SU(3) symmetry breaking pattern generates the
'missing' state in this particular sector by promoting the weak
attraction of the 27-plet contribution in (\ref{8-10-decom}).

\begin{figure}[t]
\begin{center}
\includegraphics[width=9.0cm,clip=true]{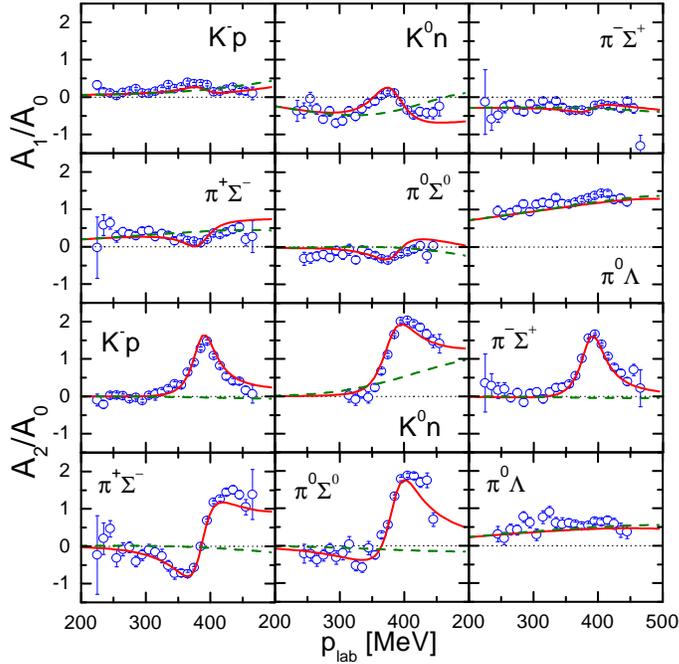}
\end{center}
\caption{Coefficients $A_1$ and $A_2$ for the $K^-p\to \pi^0
\Lambda$, $K^-p\to \pi^\mp \Sigma^\pm$ and $K^-p\to \pi^0 \Sigma$
differential cross sections\cite{mast-pio,bangerter-piS}. The
solid lines are the result of
the $\chi$-BS(3) approach with inclusion of the d-wave resonances.
The dashed lines show the effect of switching off d-wave
contributions.} \label{fig:a}
\end{figure}

\section{Chiral correction terms and scattering data}

The present data set for antikaon-nucleon scattering leaves much
room for different theoretical extrapolations to subthreshold
energies \cite{ADMartin,martsakit,kim,sakit,gopal,oades,Juelich:2,dalitz,Kaiser,Ramos}.
As a consequence the subthreshold $\bar K N$ scattering amplitudes
of different analyses may differ by as much as a factor of two
\cite{Kaiser,Ramos} in the region of the $\Lambda (1405)$
resonance. Thus it is of crucial importance to apply effective
field theory methods in order to control the uncertainties. In
particular constraints from crossing symmetry and chiral symmetry
should be taken into account.

Since the accuracy of the data improves dramatically as the energy
increases it is desirable to incorporate contributions from higher
partial waves into the analysis. Important information on the
p-wave dynamics is provided by angular distributions for the
inelastic $K^-p$ reactions. The available data are represented in
terms of coefficients $A_n$ characterizing the differential cross
section $d\sigma(\cos \theta , \sqrt{s}\,) $ as functions of the
center of mass scattering angle $\theta $ and the total energy
$\sqrt{s}$:
\begin{eqnarray}
\frac{d\sigma (\sqrt{s}, \cos \theta )}{d\cos \theta }  &=&
\sum_{n=0}^\infty A_n(\sqrt{s}\,)\,P_n(\cos \theta ) \,.
\label{a-b-def}
\end{eqnarray}
In Fig.~\ref{fig:a} we compare the empirical ratios $A_1/A_0$ and
$A_2/A_0$ with the results of the $\chi$-BS(3) approach
carried out to chiral order $Q^3$ \cite{LK02}. A large
$A_1/A_0$ ratio is found only in the $K^-p\to \pi^0 \Lambda$
channel demonstrating the importance of p-wave effects in the
isospin one channel. Note the sizeable
p-wave contributions at somewhat larger momenta seen in the
charge-exchange reaction of Fig.~\ref{fig:a}.
The dashed lines of Fig.~\ref{fig:a}, which
are obtained when switching off d-wave contributions, illustrate
the importance of the $\Lambda(1520)$ resonance for the angular
distributions in the isospin zero channel. In \cite{LK02}
the d-wave resonances were not yet generated dynamically rather were
introduced as elementary fields. To improve on this it would be necessary to
extend the analysis \cite{LK02} by incorporating further inelastic channels, like the
meson baryon-decuplet channels. Also it would be useful to incorporate the
loop-correction effects which give contributions to the effective interaction kernel at chiral order
$Q^3$. The latter were not yet considered in \cite{LK02} based on the expectation that they should
be suppressed in the large-$N_c$ limit. The importance of the results \cite{LK02} lies in the achievement
that for the first time a simultaneous and quantitative description of the low-energy pion-, kaon- and
antikaon-nucleon scattering data based on chiral coupled-channel dynamics was obtained.
Moreover that analysis demonstrated that all considered counter terms determined by
the scattering data have natural size. Though SU(3) symmetry breaking effects are important
to achieve a quantitative description of the data set they are small and justify the
application of the chiral SU(3) Lagrangian.

\section{Self consistent strangeness propagation in cold nuclear matter}

Once the microscopic interaction of the Goldstone bosons with the
constituents of nuclear matter is understood one may study the
properties of Goldstone bosons in nuclear matter. Not only from an
experimental  but also from a theoretical point of view the pions
and kaons, the lightest excitation of the QCD vacuum with masses
of $140$ MeV and $495$ MeV respectively, are outstanding probes
for exciting many-body dynamics. The Goldstone bosons are of
particular interest since their in-medium properties reflect the
structure of the nuclear many-body ground state. For example at
high baryon densities one expects the chiral symmetry to be
restored. One therefore anticipates that the Goldstone bosons
change their properties substantially as one compresses nuclear
matter. A possible consequence of a significantly reduced effective
$K^-$ mass suggested first by Kaplan and Nelson \cite{Kaplan:Nelson} could be that kaons
condense in the interior of neutron stars
\cite{K:condensation:1,K:condensation:2,Muto02,K:condensation:3}.
A complementary suggestion was recently put forward
by Yamazaki and Akaishi  \cite{Yamazaki}
that the attractive antikaon-proton force is sufficiently strong to form
high-density few-particle clusters. In this review we do not touch neither of the two
exciting topics. To address properties of antikaons at high-density
or localized strangeness systems is beyond our present scope. Our goal is to
derive the properties of antikaons in nuclear matter densities not too large, say
up to twice nuclear matter saturation density, from the microscopic interaction
of antikaons with the nucleons. This is relevant for the description of kaonic atom data
as well as for the understanding of subthreshold antikaon production in heavy ion collisions
as studied in detail at GSI \cite{Senger}.

Even though in the SU(3) limit of QCD with degenerate current
quark masses $m_u=m_d=m_s$ the pions and kaons have identical
properties with respect to the strong interactions, they provide
very different means to explore the nuclear many-body system. This
is because the SU(3) symmetry is explicitly broken  by a nuclear
matter state with strangeness density zero, a typical property of
matter produced in the laboratory. A pion, if inserted into
isospin degenerate nuclear matter, probes rather directly the
spontaneously broken or possibly restored chiral SU(2) symmetry.
A kaon, propagating in strangeness free nuclear matter, looses its
Goldstone boson character since the matter by itself explicitly
breaks the SU(3) symmetry. It is subject to three different
phenomena: the spontaneously broken chiral SU(3) symmetry, the
explicit symmetry breaking of the small current quark masses and
the explicit symmetry breaking of the nuclear matter bulk. The
various effects are illustrated by recalling the effective pion
and kaon masses in a dilute isospin symmetric nuclear matter gas.
The low-density theorem \cite{dover,njl-lutz} predicts mass
changes $\Delta m_\Phi^2$ for any meson $\Phi $ in terms of its
isospin averaged s-wave meson-nucleon scattering length $a_{\Phi
N}$
\begin{eqnarray}
\Delta m_\Phi^2 =-4\,\pi \left(1+\frac{m_\Phi}{m_N}\right) a_{\Phi
N}\,\rho + {\mathcal O} \left( \rho^{4/3} \right) \label{LDT}
\end{eqnarray}
where $\rho $ denotes the nuclear density. According to the above
arguments one expects that the pion-nucleon scattering length
$a_{\pi N} \propto m_\pi^2$ must vanish in the chiral SU(2)
limit since isospin symmetric nuclear matter conserves the
Goldstone boson character of the pions at least at small
densities. On the other hand, kaons loose their Goldstone boson
properties in strangeness-free matter and therefore one expects $a_{K N} \propto m_K$ and in
particular $a_{K^- N} \neq a_{K^+ N}$. This is demonstrated by the
Weinberg-Tomozawa theorem (see (\ref{WT-term})) which predicts the s-wave scattering
length in terms of the chiral order parameter $f \simeq 90$
MeV :
\begin{eqnarray}
a_{\pi N} = 0 + {\mathcal O} \left( m_\pi^2\right) \,, \qquad
a_{K^\pm N} = \mp \frac{m_K}{4\pi \,f^2}+ {\mathcal O} \left(
m_K^2\right) \,. \label{WT-theorem}
\end{eqnarray}
In the pion sector the Weinberg-Tomozawa theorem
(\ref{WT-theorem}) is beautifully confirmed by the smallness of
the empirical isospin averaged pion-nucleon scattering length
$a_{\pi N} \simeq -0.01$ fm. In the kaon sector the
Weinberg-Tomozawa theorem misses the empirical scattering $K^+$
nucleon scattering length $a_{K^+ N} \simeq - 0.3 $ fm by about a
factor of three. Even more spectacular is the disagreement of the
Weinberg-Tomozawa term in the $K^-$ case where (\ref{WT-theorem})
predicts $a_{K^- N} \simeq + 0.9 $ fm while the empirical $K^-$
nucleon scattering length is about $a_{K^- N} \simeq (- 0.6
+i\,1.1 ) $ fm. Whereas  in conjunction with the low-density
theorem the Weinberg-Tomozawa theorem predicts a decreased
effective $K^-$ mass, the empirical scattering length
unambiguously states that there must be repulsion in the $K^-$
channel at least at very small nuclear densities.

\begin{figure}[t]
\begin{center}
\includegraphics[width=11.5cm,clip=true]{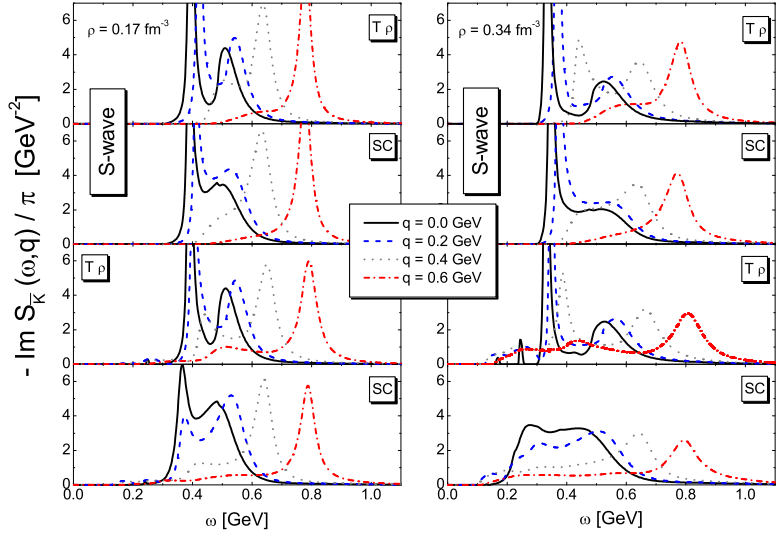}
\end{center}
\vspace*{-0.3cm}
\caption{Antikaon spectral function as a function of antikaon
energy $\omega$ and momentum $\vec q$. The labels 'T$\rho$' and 'SC'
refer to calculations obtained in terms of free-space and
in-medium $\bar K N$ amplitudes respectively.
The first two rows give the results with only s-wave interactions
and the last two rows with all s-, p- and d-wave contributions.}
\label{fig:kaon}
\end{figure}

\begin{figure}[t]
\begin{center}
\includegraphics[width=8cm,clip=true]{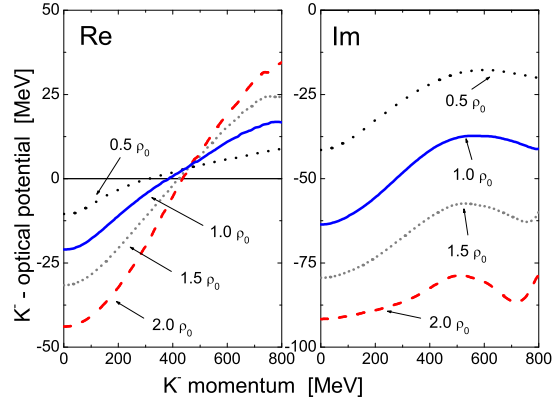}
\end{center}
\vspace*{-0.3cm}
\caption{Antikaon optical potential $V_{\rm opt.} ({\vec q}\,; \rho )$
for isospin symmetric nuclear matter densities. Results are shown for
$0.5\,\rho_0$, $1.0\,\rho_0$, $1.5\,\rho_0$ and $2.0\,\rho_0$ where
$\rho_0 = 0.17$ fm$^{-3}$.}
\label{fig:kaon-optical}
\end{figure}

In nuclear matter there exist multiple modes with quantum numbers
of the $K^-$ resulting from the coupling of the various hyperon
states to a nucleon-hole state \cite{KVK95}. As a consequence the
$K^-$ spectral function shows a rather complex structure as a
function of baryon density, kaon energy and momentum. This is
illustrated by recalling the low-density theorem as applied for
the energy dependence of the kaon self energy $\Pi_{\bar K}(\omega
, \rho)$.  At zero antikaon momentum the latter,
\begin{eqnarray}
\Pi_{\bar K}(\omega , \rho) = - 4\,\pi \left( 1+ \frac{\omega
}{m_N}\right) \,f^{\rm s-wave}_{\bar K N}(m_N + \omega )\,\rho
+{\mathcal O} \left(  \rho^{4/3}\right) \,, \label{LDT-energy}
\end{eqnarray}
is determined by the s-wave kaon-nucleon scattering amplitude
$f^{\rm s-wave}_{\bar K N}(\sqrt{s})$ (see e.g. \cite{njl-lutz}).
A pole contribution to $\Pi_{\bar K}(\omega , \rho)$ from a hyperon state with mass
$m_H$, if sufficiently strong, may lead to a $K^-$ like state of
approximate energy $ m_H-m_N$. Most important are the $\Lambda
(1405)$ s-wave resonance and the $\Sigma(1385)$ p-wave resonance.
The realistic evaluation of the antikaon self energy
in nuclear matter requires a self-consistent scheme \cite{ml-sp}. In particular
the feedback effect of an attractive antikaon spectral function on
the antikaon-nucleon scattering process was found to be important
for the $\Lambda(1405)$ resonance structure in nuclear matter. This
effect is easily understood from the strong dependence of the
antikaon-nucleon scattering amplitude on the kaon mass as demonstrated in
Fig. \ref{fig4}. It has striking consequences for the properties of the
$\Lambda(1405)$ resonance in nuclear matter. Contrary to naive expectations \cite{Koch}
that the effective resonance mass is pushed up to larger values due to the Pauli blocking effect,
an attractive mass shift results as a consequence of self consistency
\cite{ml-sp,Lutz:Korpa}. Ramos and Oset \cite{ramossp} confirmed this result qualitatively by a
calculation applying their phenomenological model \cite{Ramos}. The
quantitative comparison of the original calculation \cite{ml-sp}, which was based on the
model of the Munich group \cite{Kaiser}, is hampered by the facts that first the
subthreshold amplitudes of \cite{Kaiser,LK02} and \cite{Ramos} differ
significantly and second the computation in \cite{ramossp} relies on an additional
approximation that amounts to neglecting the strong momentum dependence of the
antikaon self energy (see \cite{LH02,Cieply}). We refrain here from a
comparison with the results of Tolos et al. \cite{Tolos}. Their many-body
computation includes higher partial-wave contributions but self consistency
was implemented relying on a quasi-particle approximation. Moreover, the applied
meson-exchange model \cite{Juelich:2} was confronted so far with total cross
sections data only. Thus, it remains unclear whether it describes the dynamics of higher
partial waves correctly. A precise understanding of all these issues
is required for a microscopic description of kaonic atoms \cite{Florkowski,Cieply}.

\begin{figure}[t]
\begin{center}
\includegraphics[width=11.5cm,clip=true]{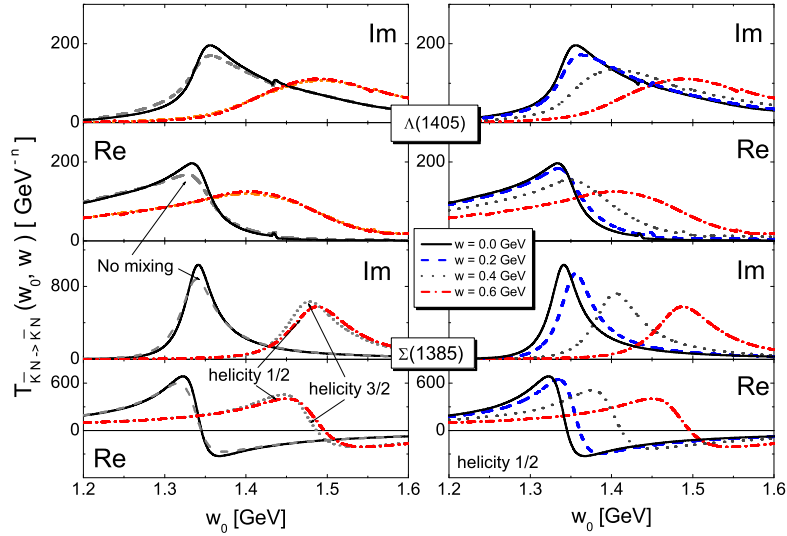}
\end{center}
\vspace*{-0.3cm}
\caption{Hyperon resonance spectral functions for the $\Lambda(1405)$ and
$\Sigma (1385)$ as a function of the hyperon energy $w_0$ and momentum $\vec w$
for isospin symmetric nuclear matter. It is shown the
s-wave I=0 and p-wave I=1 partial wave amplitudes at $\rho_n+\rho_p= 0.17$ fm$^{-3}$.
The left hand panels demonstrate the small effect of switching off the mixing of the partial
wave amplitudes (dashed versus full lines) and the splitting of the 4 spin 3/2 states into
helicity one half and three half modes (dotted versus dashed-dotted lines).}
\label{fig:hyperon-spec-1}
\end{figure}

\begin{figure}[t]
\begin{center}
\includegraphics[width=11.5cm,clip=true]{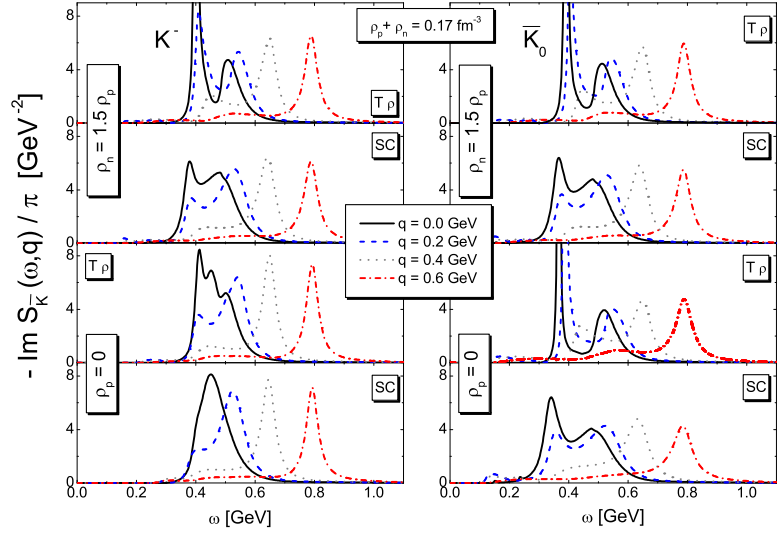}
\end{center}
\vspace*{-0.3cm}
\caption{Antikaon spectral functions as a function of antikaon
energy $\omega$ and momentum $q$ in asymmetric nuclear matter
with $\rho_p\!+\!\rho_n = 0.17$ fm$^{-3}$. The labels 'T$\rho$' and 'SC'
refer to calculations obtained in terms of free-space and self consistent
in-medium $\bar K N$ amplitudes respectively.
The first collum gives the result for $K^{-}$, the second for the
$\bar K_0$ at $\rho_n=1.5 \,\rho_p$ and $\rho_p=0$.}
\label{fig:kaon-sp-as}
\end{figure}

\begin{figure}[t]
\begin{center}
\includegraphics[width=11.5cm,clip=true]{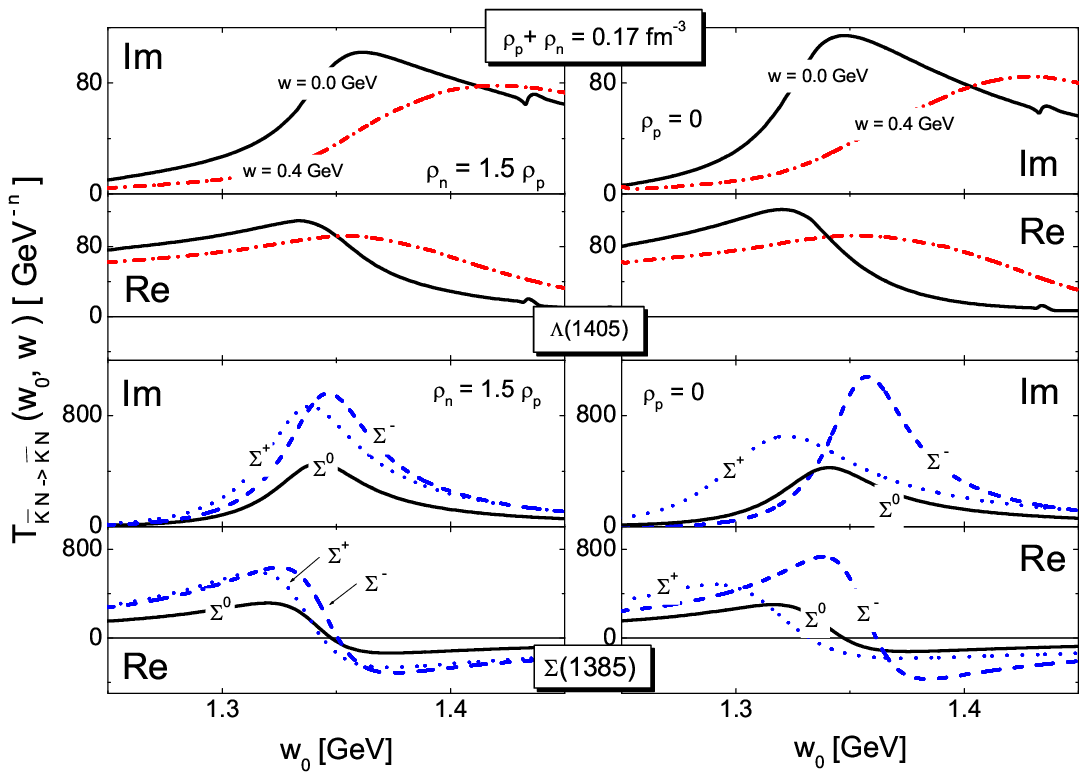}
\end{center}
\vspace*{-0.3cm}
\caption{$\Lambda(1405)$ and $\Sigma (1385)$ hyperon resonance propagators as a
function of hyperon energy $w_0$ and momentum $\vec w$ for asymmetric nuclear matter.
It is shown the s-wave $K^- p \to K^- p$ (for $\vec w= 0$ MeV and $\vec w= 400$ MeV) and
p-wave  $\bar K^0 n \to \bar K^0 n$, $\bar K^0 p \to \bar K^0 p$ and
$K^- n \to K^- n$ reaction amplitudes (for $\vec w= 0$ MeV) at $\rho_n+\rho_p= 0.17$ fm$^{-3}$.}
\label{fig:hyperon-sp-as}
\end{figure}

In Fig. \ref{fig:kaon} the antikaon spectral function
evaluated at symmetric nuclear densities $\rho_0$ and $2\,\rho_0$ according
to various approximation strategies is shown \cite{Lutz:Korpa}. The
results are based on antikaon-nucleon scattering amplitudes
obtained within the chiral coupled-channel effective field theory
\cite{LK02}, where s-, p-and d-wave contributions were considered.
The many-body computation \cite{Lutz:Korpa} was performed in a self
consistent manner respecting covariance manifestly. In the first and
third rows the antikaon self energy is computed in terms of the
free-space scattering amplitudes only. Here the first row gives
the result with only s-wave contributions and the third row
includes all s-, p- and d-wave contributions. The second and fourth
rows give results obtained in the self
consistent approach where the full result
of the last row includes all partial waves and the results in the
second row follow with s-wave contributions only. In all cases a
self consistent evaluation of the spectral function
leads to dramatic changes in the spectral function as compared to
a calculation which is based on the free-space scattering
amplitudes only. Moreover, as emphasized in the discussion of the
$\Lambda(1405)$ resonance properties, the effects of higher
partial waves are not negligible. This was anticipated first in
\cite{LK00,Florkowski}. As is evident upon comparing the
first and third rows of Fig. \ref{fig:kaon} the p- and d-wave
contributions add quite significant attraction for small energies
and large momenta. At twice nuclear saturation density
most striking is the considerable support of the spectral function at
small energies. That reflects in part the coupling of the antikaon
to the $\Lambda(1115)$ and $\Sigma (1185)$ nucleon-hole states.

The non-trivial dynamics implied by the presence of the hyperon resonances
reflects itself in a complicated behavior of the antikaon self energy, $\Pi_{\bar K}( \omega
, {\vec q}\,; \rho )$. This is illustrated by the antikaon nuclear optical potential
$V_{\rm opt.} (\vec q \,; \rho )$, which may be defined by
\begin{eqnarray}
2\,E_K({\vec q}\,)\, V_{\rm opt.} ({\vec q}\, ; \rho) =\Pi_{\bar K}( \omega
=E_K({\vec q}\,), {\vec q}\,; \rho ) \;, \qquad E_K({\vec q}\, )=\sqrt{m_K^2+{\vec q}\,^2 }\,.
\end{eqnarray}
In Fig. \ref{fig:kaon-optical} the result for the optical potential is presented
as a function of the antikaon momentum $\vec q $ and the nuclear matter density.
The real part of the optical potential exhibits rather moderate
attraction of about 20 MeV at nuclear saturation density and $\vec q= 0$ MeV. The attraction
is further diminished and even turns into repulsion as the antikaon momentum increases.
On the other hand the figure shows a rather strong absorptive
part of the optical potential. This agrees qualitatively with computations
based on self consistent s-wave dynamics \cite{Florkowski} but is in striking disagreement
with mean-field calculations \cite{Schaffner:Bondorf:Mishustin} that predict considerable
more attraction in the antikaon optical potential. The large attraction in the antikaon
spectral function of Fig. \ref{fig:kaon} is consistent with the
moderate attraction in the optical potential of Fig. \ref{fig:kaon-optical}. It merely
reflects the strong energy dependence of the kaon self energy induced by the
$\Lambda (1405) $ and $\Sigma (1385 )$ resonances. Such important energy variations are
missed in a mean-field approach. Hence, a proper treatment
of the pertinent many-body effects is required.

In Fig. \ref{fig:hyperon-spec-1} the propagation
properties of the $\Lambda(1405)$ and $\Sigma (1385)$ resonances as they move
with finite three momentum $\vec w$ in isospin symmetric nuclear matter are studied.
The peak positions in the imaginary parts of the amplitudes follow in general the naive
expectation $\sqrt{(m_Y^*)^2+\vec w^2}$ with the hyperon effective mass $m_Y^*$ defined
at $\vec w=0$. However, a systematic increase of the decay widths as
$\vec w$ increases is observed. This is easily understood since the effective in-medium
$Y^* N \to Y^* N$ amplitude which is responsible for the broadening allows
for additional inelasticity as the hyperon momentum $\vec w$ increases.
An interesting phenomenon illustrated in Fig. \ref{fig:hyperon-spec-1}
is the in-medium induced mixing of partial wave amplitudes with
different quantum numbers $J^P$. At vanishing three
momentum of the meson-baryon state $\vec w = 0$, all partial-wave
amplitudes decouple. In this case the system enjoys a three-dimensional rotational symmetry since
there is no three-vector available to select a particular direction.
However, once the meson-baryon pair is moving relative to the nuclear matter bulk with
$\vec w \neq 0$ there are two separate channels for a given isospin only. The system is invariant under
a subgroup of rotations only, namely those for which the rotational vector is aligned with
the hyperon momentum $\vec w$. The three-dimensional rotational symmetry is reduced and therefore
the total angular momentum $J$ is no longer a conserved quantum number. However, the total angular
momentum projection onto the $\vec w$ direction remains a conserved quantum number. The latter
defines the helicity of a hypothetical s-channel particle exchange and therefore the in-medium scattering
amplitudes decouple into an infinite tower of helicity amplitudes. Each helicity amplitude probes a well
defined infinite set of partial wave amplitudes. For instance, in the helicity one-half space,
all considered partial wave amplitudes $S_{I1}$, $P_{I1}$, $P_{I3}$ and $D_{I3}$ couple for given
isospin channel $I$. Note that
reflection symmetry implies that only the absolute
value of the helicity matters here. It is evident that all partial-wave amplitudes
contribute to this channel simply because any state with a given angular momentum $L$ has a
component where $\vec L \cdot \vec w $ vanishes and therefore the helicity is carried by the
nucleon spin. In the second channel, the helicity three-half space, only the partial wave
amplitudes $P_{I3}$ and $D_{I3}$ couple. In a computation
that considered all partial waves the helicity three half term would require all partial wave
amplitudes except the $S_{I1}$ and $P_{I1}$ waves with $J=1/2$. The fact that the
$J={\textstyle{3\over 2}}$ amplitudes $P_{I3}$ and $D_{I3}$ affect both considered helicity spaces,
is not surprising, because for those
states one would expect the nuclear medium to lift the degeneracy of the four spin modes. This
is completely analogous to the longitudinal and transverse modes of
vector mesons, which bifurcate in nuclear matter.

The splitting of the four $\Sigma (1385)$ modes is demonstrated in the left hand
panel of Fig. \ref{fig:hyperon-spec-1}. The helicity one and three half modes are
shifted by about 5 MeV for $\vec w = 600$ MeV. This is a small effect and therefore
it is justified to neglect the coupling of the partial waves for
that density to good accuracy. Indeed a run where all off diagonal loop functions
that couple waves of different angular momentum or parity are set to zero gives
results that are almost indistinguishable to those obtained in the full scheme. This
is demonstrated by the dashed lines in the left hand panel of Fig. \ref{fig:hyperon-spec-1}.
At $\vec w =0$ the results of the two computations, dashed and solid lines are quite close.
This finding has a simple interpretation. One expects sizeable effects from the in-medium
mixing of the partial-wave amplitudes only if two partial wave amplitudes that mix show both
significant strength at a given energy $w_0$. This is not
the case here. The dominant partial waves $S_{01}$ and $P_{13}$ decouple because they
carry different total isospin. Therefore it is natural to obtain small mixing effects.

A discussion of results \cite{Lutz:Korpa,LH02,Korpa:Lutz:03} obtained for asymmetric nuclear matter follows.
Two cases are considered here both with $\rho_n+\rho_p= 0.17$ fm$^{-3}$.
In Fig. \ref{fig:kaon-sp-as} results for $\rho_n= 1.5\,\rho_p$, which corresponds to
the condition met in the interior of lead, and for $\rho_p=0$, which describes neutron matter,
are shown. The asymmetry breaks up the isospin doublet $(K^-, \bar K_0)$ leading to
distinct spectral functions for the charged and neutral antikaons. In both cases
a significant effect from the self consistency is found at small antikaon momenta. This
is illustrated by comparing the entries of Fig. \ref{fig:kaon-sp-as} labeled
with 'T$\rho$' and 'SC'. Whereas the effect of the asymmetry
is surprisingly small for the lead scenario, the spectral functions of the
charged and neutral antikaons differ strongly in neutron matter with $\rho_p=0$.
In Fig. \ref{fig:hyperon-sp-as} the corresponding properties of the
hyperon resonances are shown. Again, like one observed for the
isospin doublet $(K^-,\bar K^0)$ the isospin asymmetry of the
matter breaks up the isospin triplet state $\Sigma(1385)$
introducing a medium-induced splitting pattern. The
$\Lambda(1405)$ resonance is presented in terms of the in-medium
$K^- p \to K^- p$ s-wave amplitude, the neutral and charged
$\Sigma(1385)$ states by p-wave amplitudes $\bar K^0 n \to \bar
K^0 n$, $\bar K^0 p \to \bar K^0 p$ and $K^- n \to K^- n$. In free
space the latter amplitudes are determined by the isospin one
amplitude in case of the charged hyperon states but by the mean of
the two isospin amplitudes in case of the neutral hyperon state.
Thus, the isospin one component, the $\Sigma (1385)$ couples to in
free space, is smaller by a factor of two for the neutral
amplitude $\bar K^0 n \to \bar K^0 n$ as compared to the
amplitudes describing the charged hyperon states. This reflects
itself in amplitudes for the $\Sigma^0(1385)$ in Fig.
\ref{fig:hyperon-sp-as} that are typically smaller by a factor two
as compared to the amplitudes of the $\Sigma^\pm(1385)$. Whereas
the $\Lambda(1405)$ resonance is not affected much by the
asymmetry, the splitting of the three $\Sigma (1385)$ states shows
a strong dependence on the asymmetry. For neutron matter with
$\rho_n = 0.17$ fm$^{-3}$  a mass difference of about 30
MeV for the charged states is found.

\section{Summary}

In this talk we reported on recent progress in the understanding
of baryon resonances based on chiral-coupled channel dynamics. An
introduction to an effective field theory formulation of
chiral coupled-channel dynamics was given. Leading order results predict the
existence of s- and d-wave baryon resonances with a spectrum
remarkably close to the empirical pattern without any adjustable
parameters. The formation of resonances is a consequence of the
chiral SU(3) symmetry of QCD, i.e. in an effective field theory,
that was based on the chiral SU(2) symmetry only, no resonances
would be formed. Realistic scattering amplitudes
that are consistent with empirical differential cross sections
can be obtained after including chiral corrections terms
systematically.

As a further application of chiral coupled-channel dynamics
results for antikaon and hyperon resonance propagation in cold
nuclear matter were presented. Based on scattering amplitudes obtained
within the chiral coupled-channel effective field theory a self consistent
density summation scheme can be performed that respects covariance manifestly.
As a consequence the spectral function of the antikaon shows a
strong momentum and density dependence in isospin symmetric and asymmetric matter.
For the $\Lambda(1405)$ and $\Sigma(1385)$ resonances attractive mass shifts are
predicted.

%


\begin{thebibliography}{99}


\bibitem{ADMartin}
A.D.~Martin, Nucl. Phys. {\bf B 179} (1981) 33.

\bibitem{dover} C.D.~Dover, J.~H\"ufner and R.H.~Lemmer,
Ann. Phys. {\bf 66} (1971) 248.

\bibitem{njl-lutz}
M.~Lutz, A. Steiner and W. Weise, Nucl. Phys. {\bf A 574} (1994)
755.

\bibitem{ml-sp}
M. Lutz, Phys. Lett. {\bf B 426} (1998) 12; M.F.M. Lutz, in {\it
Proc. Workshop on Astro-Hadron Physics}, Seoul, Korea, October,
1997, World Scientific 1999.

\bibitem{LK02}
M.F.M. Lutz and E.E. Kolomeitsev, Nucl. Phys. {\bf A 700} (2002) 193.

\bibitem{Lutz:Korpa}
M.F.M. Lutz and C.L. Korpa, Nucl. Phys. {\bf A 700} (2002) 309.

\bibitem{Korpa:Lutz:03}
C.L. Korpa and M.F.M. Lutz, Heavy Ion Phys. {\bf 17} (2003) 341.

\bibitem{Florkowski}
M.F.M. Lutz and W. Florkowski, Act. Phys. Pol. {\bf 31} (2000) 2567.

\bibitem{Cieply}
A. Cieply, E. Friedman, A. Gal and J. Mares, Nucl. Phys. {\bf A 696} (2001) 173.

\bibitem{LK01}
M.F.M. Lutz und E.E. Kolomeitsev, Found. Phys. {\bf 31} (2001)
1671.

\bibitem{LWF02}
M.F.M. Lutz, Gy. Wolf and B. Friman, Nucl. Phys. {\bf A 706}
(2002) 431.

\bibitem{LH02}
M.F.M. Lutz, GSI-Habil-2002-1.

\bibitem{Granada} C. Garc\'\i a-Recio, M.F.M. Lutz and J.
Nieves, Phys. Lett. {\bf B 582} (2004) 49.

\bibitem{KL03}
E.E. Kolometsev and M.F.M. Lutz, Phys. Lett. {\bf B 585} (2004) 243.

\bibitem{LK03}
M.F.M. Lutz and E.E. Kolomeitsev,
Nucl. Phys. {\bf A 730} (2004) 392.

\bibitem{KL03-2}
E.E. Kolomeitsev and M.F.M. Lutz,
Phys. Lett. {\bf B 582} (2004) 39.

\bibitem{LK03-2}
M.F.M. Lutz and E.E. Kolomeitsev,
Nucl. Phys. {\bf A 730} (2004) 110.

\bibitem{HL03}
J. Hofmann and M.F.M. Lutz,
Nucl.Phys. {\bf A 733} (2004) 142.

\bibitem{Wyld}
H.W. Wyld, Phys. Rev. {\bf 155} (1967) 1649.

\bibitem{Dalitz} R.H. Dalitz, T.C. Wong and G. Rajasekaran,
Phys. Rev. {\bf 153} (1967) 1617.

\bibitem{Ball}
J.S. Ball and W.R. Frazer, Phys. Rev. Lett. {\bf 7} (1961) 204.

\bibitem{Rajasekaran}
G. Rajasekaran, Phys. Rev. {\bf 5} (1972) 610.

\bibitem{Wyld2}
R.K. Logan and H.W. Wyld, Phys. Rev. {\bf 158} (1967) 1467.

\bibitem{sw88} P.B.~Siegel and W.~Weise,
Phys. Rev. {\bf C 38} (1988) 2221.

\bibitem{LK00}
M.F.M. Lutz and E. E. Kolomeitsev, Proc. of Int. Workshop XXVIII
on Gross Properties of Nuclei and Nuclear Excitations, Hirschegg,
Austria, January 16-22, 2000.

\bibitem{Krause}
A.~Krause, Helv. Phys. Acta {\bf 63} (1990) 3.

\bibitem{book:Weinberg}
S. Weinberg, {\it The quantum theory of fields}, Vol. II,
University Press, Cambridge (1996).

\bibitem{Lutz00}
M.F.M. Lutz, Nucl. Phys. {\bf 677} (2000) 241.

\bibitem{NA00}
J. Nieves and E. Ruiz Arriola, Nucl. Phys. {\bf A 679} (2000) 57.

\bibitem{Wein-Tomo}
S. Weinberg, Phys. Rev. Lett. {\bf 17} (1966) 616;\\
Y. Tomozawa, Nuov. Cim. {\bf A 46} (1966) 707.

\bibitem{ksw95}
N.~Kaiser, P.B.~Siegel and W.~Weise,
Nucl. Phys. {\bf A 594} (1995) 325.

\bibitem{grnpi}
J. Nieves and E. Ruiz Arriola, Phys. Rev. {\bf D 63}, (2001) 076001.

\bibitem{grkl} C. Garc\'\i a-Recio,
J. Nieves, E. Ruiz Arriola and  M. J. Vicente-Vacas,
Phys. Rev. {\bf D 67} (2003) 076009.

\bibitem{Oset-prl} A. Ramos, E. Oset and C. Bennhold,
Phys. Rev. Lett. {\bf 89} (2002) 252001.

\bibitem{Oset-plb}
E. Oset, A. Ramos, C. Bennhold,  Phys. Lett. {\bf B 527} (2002) 99.

\bibitem{Jido03}
D. Jido et al.,  Nucl. Phys. {\bf A 725} (2003) 181.

\bibitem{Schat}
C.L. Schat, J.L. Goity and N.N. Scoccola, Phys. Rev. Lett. {\bf 88} (2002) 102002.

\bibitem{lattice:1}
F. Karsch, Nucl. Phys. B (Proc. Suppl.),83-84 (2000) 14.

\bibitem{lattice:2}
G. Boyd, et al. Phys. Lett. {\bf B 349} (1995) 170.

\bibitem{lattice:3}
T. Hatsuda, hep-ph/0104139.

\bibitem{dgr} D. G. Richards, Proc. of `NSTAR 2002', Pittsburgh,
October 2002, and references therein.

\bibitem{Hoehler:speed}
G. H\"ohler, $\pi N$ NewsLetter {\bf 9} (1993) 1.

\bibitem{martsakit}
B.R.~Martin and M.~Sakit, Phys. Rev. {\bf 183} (1969) 1352.

\bibitem{kim}
J.K.~Kim, Phys. Rev. Lett. {\bf 14} (1965) 29.

\bibitem{sakit}
M.~Sakit et al., Phys. Rev. {\bf 139} (1965) B179.

\bibitem{gopal}
G.P. Gopal et al. Nucl. Phys. {\bf 119} (1977) 362.

\bibitem{oades}
G.~C.~Oades, in Proc. of {\it Int. Workshop on Low and
Intermediate-Energy Kaon-Nucleon Physics}, Rome, Italy, 24-28
March 1980, (eds. E. Ferrari, G. Violini), D. Reidel Publishing,
Dordrecht (1980).

\bibitem{Juelich:2}
A. M\"uller-Groling, K. Holinde and J. Speth, Nucl. Phys. {\bf A 513} (1990) 557.

\bibitem{dalitz}
R.H. Dalitz and A. Deloff, J. Phys. {\bf G 17} (1991) 289.

\bibitem{Kaiser}
N.~Kaiser, P.B.~Siegel and W.~Weise,
Nucl. Phys. {\bf A 594} (1995) 325; \\
N.~Kaiser, T.~Waas and W.~Weise, Nucl. Phys. {\bf A 612} (1997) 297.

\bibitem{Ramos}
E.~Oset and A.~Ramos, Nucl. Phys. {\bf A 635} (1998) 99.

\bibitem{mast-pio}
T.S.~Mast et al., Phys. Rev. {\bf D~11} (1975) 3078.

\bibitem{bangerter-piS}
R.O.~Bangerter et al., Phys. Rev.\ {\bf D~23} (1981) 1484.

\bibitem{Kaplan:Nelson}
D.B. Kaplan and A.E. Nelson, Phys. Lett. {\bf B 175} (1986) 57.

\bibitem{K:condensation:1}
G.E. Brown and H.A. Bethe, Astrophys. Jour. {\bf 423} (1994) 659.

\bibitem{K:condensation:2}
G.Q. Li, C.-H.Lee, G.E. Brown, {Phys. Rev. Lett.} {\bf 79} (1997) 5214; Nucl. Phys. {\bf A 625} (1997) 372.

\bibitem{Muto02}
T. Muto and T. Tatsumi, Phys. Lett. {\bf B 283} (1992) 165;
T. Muto, Nucl. Phys. {\bf A 697} (2002) 225.

\bibitem{K:condensation:3}
E.E. Kolomeitsev and D.N. Voskresensky, Phys. Rev. {\bf C 68} (2003) 015803.

\bibitem{Yamazaki}
A. Dote, H. Horiuchi, Y. Akaishi and T. Yamazaki, nucl-th/0207085 , nucl-th/0309062;

\bibitem{Senger}
F.~Laue, Ch.~Sturm et al., Phys. Rev. Lett. {\bf 82} (1999) 1640.

\bibitem{KVK95}
E.E. Kolomeitsev, D.N. Voskresensky and B.
K\"ampfer, Nucl. Phys. {\bf A 588} (1995) 889.

\bibitem{Koch}
V. Koch, Phys. Lett. {\bf B 337} (1994) 7.

\bibitem{ramossp}
A. Ramos and E. Oset, Nucl. Phys. {\bf A 671} (2000) 481.

\bibitem{Tolos}
L. Tolos, A. Ramos, A. Polls and T.S. Kuo, Nucl. Phys. {\bf A 690} (2001) 547.

\bibitem{Schaffner:Bondorf:Mishustin}
J. Schaffner, J. Bondorf and I.N. Mishustin, Nucl. Phys. {\bf A 625} (1997) 325.



\end{thebibliography}
\end{document}